\documentclass[10pt,journal,compsoc]{IEEEtran}
\usepackage{array}

\ifCLASSOPTIONcompsoc
   \usepackage[nocompress]{cite}
\else
    \usepackage{cite}
\fi

\ifCLASSINFOpdf
 
\else
 
\fi

\usepackage{graphicx}
\usepackage{xcolor}
\usepackage{multirow}
\usepackage{hyperref}

\begin{document}

\title{AI-Assisted Authentication: State of the Art, Taxonomy and Future Roadmap}


\author{Guangyi~Zhu,
        Yasir~Al-Qaraghuli
\IEEEcompsocitemizethanks{\IEEEcompsocthanksitem Guangyi Zhu is with the School of Computer Science, University of Guelph, Ontario,
N1G 2W1, Canada.\\
E-mail: guangyi@uoguelph.ca

\IEEEcompsocthanksitem Yasir Al-Qaraghuli is with the School of Computer Science, University of Guelph, Ontario,
N1G 2W1, Canada.\\
E-mail: yalqarar@uoguelph.ca}
}


\IEEEtitleabstractindextext{%
\begin{abstract}
Artificial Intelligence (AI) has found its applications in a variety of environments ranging from data science to cybersecurity. AI helps break through the limitations of traditional algorithms and provides more efficient and flexible methods for solving problems. In this paper, we focus on the applications of artificial intelligence in authentication, which is used in a wide range of scenarios including facial recognition to access buildings, keystroke dynamics to unlock smartphones. With the emerging AI-assisted authentication schemes, our comprehensive survey provides an overall understanding on a high level, which paves the way for future research in this area. In contrast to other relevant surveys, our research is the first of its kind to focus on the roles of AI in authentication.
\end{abstract}

}

\maketitle

\IEEEdisplaynontitleabstractindextext

\IEEEpeerreviewmaketitle

\section{Introduction and Basic Concepts}\label{sec:introduction}

The traditional password-based authentication method has slowly faded out due to its inadequate security \cite{a1}.  Artificial Intelligence (AI) enriched authentication mechanisms, to enhance the security of authentication and to be compatible with as many practical scenarios as possible, AI-assisted authentication methods were leading the trend in the past few years. Integrated with machine learning, deep neural networks, AI-assisted authentication can meet a variety of needs. In this paper, we review existing research. We start by defining AI and machine learning terms. We review the state-of-the-art of AI-assisted authentication. Particularly, we present a taxonomy of AI-assisted Authentication schemes, and lastly, provide future research directions and road maps.

In the past decade, Artificial Intelligence has been on the rise. Authentication is a crucial technology to ensure the security of information systems.With time, authentication has gradually adopted various AI technologies to achieve a diverse and higher accuracy. As support to authentication methods, AI can either be a simple learning model, or a deep neural network, which accepts more complex stages of data. Assistance of AI authentication is no longer limited to traditional passwords. A variety of different authentication methods are based on image recognition such as fingerprint, facial recognition, and some are based on various human behaviour such as keystroke dynamics and mouse movement, biometrics authentication such as ECG-based authentication.

\subsection{AI, Learning and Artificial Neural Networks}

\textcolor{black}{Artificial Intelligence (AI) is a field of study that focuses on computers and machines to learn from humans. Machines are taught problem solving to be able to make decisions similar to human beings. Learning and neural networks are two main mechanisms used in AI. Learning is the process of inputting data or information to a system. This information is then used to analyze problems and solve them using the information that was inputted.}

Artificial Neural Network is a model network that learns how to make decisions similar to a human's mind, this is done by simulating the nerves in a human's body.

\subsubsection{Feed Forward and Recurrent Neural Networks}

\textcolor{black}{Neural networks can be divided to feed forward and recurrent neural networks.}
\textcolor{black}{Neural networks are inspired by biology and became the heart of deep learning. Neural network is divided into feed-forward neural networks and recurrent neural networks.}

Feed Forward Neural Networks is a type of artificial neural network. It is called feed forward because information flows in one direction, information moves from the input nodes to the output nodes without any loops or cycles. \textcolor{black}{Recurrent neural networks (RCC) loops over the network multiple times until it achieves the highest prediction accuracy. RCC uses time-series data to produce predictions based on the previous output. Such neural networks are naturally compatible with ordinal problems like voice recognition.}

\subsubsection{Supervised, Semi-Supervised and Reinforcement Learning}

Supervised learning is one of the subcategories of machine learning. It describes a method that uses labelled data to train machine learning algorithms. By learning from those labelled data, algorithms can finally classify data or predict the trend. For example, we feed a machine learning model scam emails and real emails to detect scam emails and flag them.

Semi-Supervised learning has the same function as Supervised learning. Semi-Supervised learning is the use of labelled and unlabelled data to train a model.

\textcolor{black}{Reinforcement learning (RL) is a field of machine learning that focuses on the concept of how a machine should behave in an environment to maximize prediction accuracy. What differentiates it from supervised and unsupervised learning is that reinforcement learning does not need labelled input or output and sub-optimal actions to achieve high accuracy.}

\subsubsection{Adversarial, Multi-Task, In-Situ and Ensemble Learning}

\textcolor{black}{Adversarial Learning is a machine learning technique that generates malicious information to violate the rules, in order to cause malfunction during learning and to create attacks.}

Multi-Task learning is the process where multiple models are learning simultaneously. The models must have some sort of correlation between them that will enhance the learning process by speeding it up and reducing false outputs.

In-situ learning is the process where input data is stored within the machine learning model. This process saves time and money. Instead of getting data from databases running an algorithm and updating it with the output, it is automatically integrated into the database that's within the model.

\textcolor{black}{Ensemble Learning is a machine learning scheme that allows AI to combine more than one algorithm or model to achieve high prediction performance that could not be achieved by any single model alone.}

\subsubsection{Feature Learning and Metric Learning}

Feature learning, also known as representation learning, is an approach that enables machine learning systems to extract features needed for further processing from raw data. With it, the machine learning system can learn and use these features at the same time.

Metric learning or distance metric learning is a way that can automatically build up distance metrics for a specific task. It can learn from supervised or weakly supervised data to construct metrics of the distance between data points. And then use the metrics to perform like classification, clustering.

\subsubsection{Meta Learning}

Meta Learning is named after the procedure that takes meta data from already done machine learning models. The input of the meta learning model is the meta data of another machine learning model, it helps with picking the best algorithm with the best outcomes.

\subsection{Deep Neural Networks, Machine Learning and Deep Learning}

\textcolor{black}{Deep Neural Network (DNN) is a network model built to imitate the activities of the human brain \cite{a7}.} A neural network contains an input layer, an output layer, and at least one hidden layer in between. The hidden layers use plenty of data to simulate the thinking pattern of the human brain in a complex way by building a very sophisticated mathematical model.

Machine learning (ML) is a sub-field of Artificial Intelligence (AI). By combining with computer science and mathematics, machine learning focuses on building models with algorithms. With training by data, machine learning models can predict or decide like a human brain without any explicit programming. And the accuracy of the output can gradually grow as the number of learning increases.

Deep learning, also known as Deep Machine Learning, is considered to be a premium member of machine learning, which uses deep neural networks. It can learn from a large amount of data, which is not necessary to be in a structured format. Deep learning algorithms are able to find the most important features to distinguish the samples while for general machine learning, feature vectors are needed to be built by humans manually.

Due to their ability to process large volumes of data, AI-based methods have been used in a variety of scientific and technological areas. Among these areas, one may refer to communications \cite{AI-Appl-Jour001}\cite{AI-Appl-Jour002,a4}, robotics \cite{AI-Appl-Jour003}, big data \cite{AI-Appl-Jour004}, education \cite{AI-Appl-Jour005}, medicine \cite{AI-Appl-Jour006}, urban services \cite{AI-Appl-Jour007}, etc\cite{a5}. 

They have found their applications in several security-related ares such as privacy \cite{Appl-Se-Jour001}, trust \cite{Appl-Se-Jour002,a6}, attack detection \cite{Appl-Se-Jour003,a24},  encryption \cite{Appl-Se-Jour004}, secret sharing \cite{Appl-Se-Jour005,a3} and watermarking \cite{Appl-Se-Conf001}. In this review, we focus on the applications of AI in authentication. 

\begin{figure}[htbp]
	\centering
	 \includegraphics[width=0.8\linewidth,keepaspectratio]{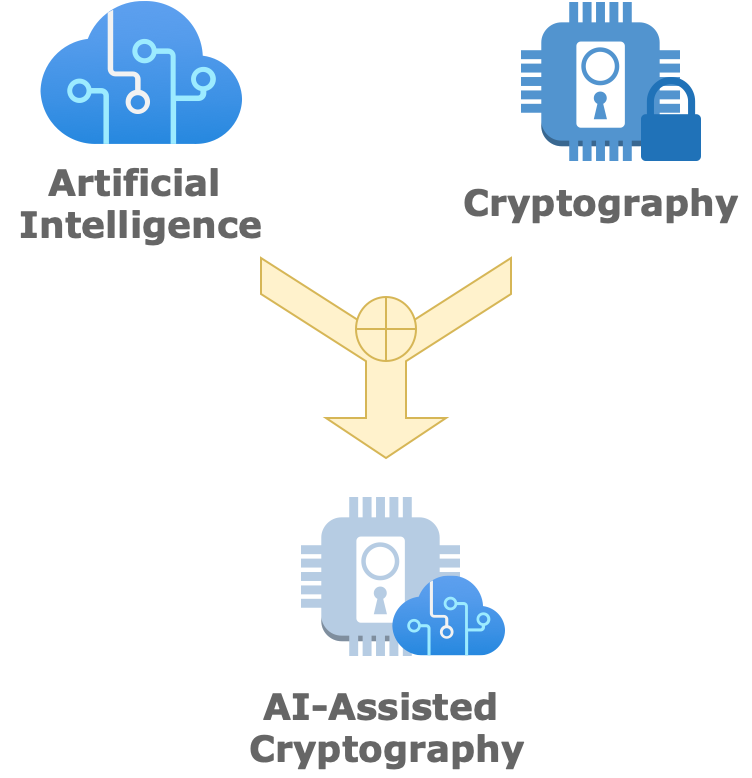}
	\caption{AI-Assisted Authentication}
	\label{AI_and_Auth}
\end{figure}

\textcolor{black}{AI is of high importance in authentication technology. Traditional authentication methods identify people based on the knowledge that they have, such as passwords. Those schemes have poor security and are extremely limited in some situations. By adopting AI, authentication can be done based on tokens that people have, biometric features that describe a specific person, even the way people behave. Moreover, AI-assisted authentication highly increased the users' experience. For instance, a smartphone can authenticate a user without being noticed. Therefore, AI has been gradually embedded into the legacy authentication process to provide high security and a satisfactory user experience. The relationship between Authentication and AI is shown in the Figure \ref{AI_and_Auth}.}

 As seen in the Figure \textcolor{black}{\ref{AI_and_Auth}}, AI-assisted authentication is a combination of Artificial Intelligence and traditional identity authentication, such as password authentication. Such techniques can use more complex information to identify an individual. It can provide higher authentication accuracy while improving security.

A comprehensive survey can pave the way for further research in this area. Numerous AI-assisted authentication methods have been proposed and implemented.

The paper will search for relevant papers and go over existing surveys about the state of the art, taxonomy and future roadmap. It has been difficult to fully summarize and analyze such surveys due to the rapidly increasing  number of research on AI-assisted authentication. Although we have already had quantities of existing surveys in this area, most of them did not cover the state of the art, taxonomy and future roadmap at the same time. So, our review is worth great value due to its completeness.

Although there are already some similar surveys that have been conducted, they are not sufficient to provide a complete and comprehensive review of AI-assisted Authentication. Like what is going to discuss in section \ref{EX-Surv}, those surveys are more likely to focus on specific aspects.  

The rest of this paper is organized as follows. Section \ref{EX-Surv} presents a review on existing surveys to highlight our motivations for this review. Section \ref{State} discusses the state-of-the-art of AI-assisted authentication. Section \ref{Tax} presents our taxonomy on the AI-based methods used in authentication. Section \ref{Road} establishes a future roadmap for research in this area. Lastly, Section \ref{Conc} concludes the paper and suggest future works.

\textcolor{black}{Figure \ref{AI_and_Auth} shows the relation between AI and Cryptography in authentication. As seen in the graph, AI and Modern Cryptography were combined. The lock icon in Cryptography represents the traditional encrypt and decrypt algorithms. After it was combined with AI, the lock icon was replaced with an AI icon, which means AI can extend the algorithms and produce more reliable and efficient ways to authenticate an entity.}

\section{Existing Surveys}\label{EX-Surv}

There are many surveys focusing on AI. However, some of them are too outdated for such a dynamic research area \cite{Surv-Conf003}\cite{Surv-Conf001}\cite{Surv-Jour008}. Some of them focus only on a branch of AI or a specific application area \cite{Surv-Jour001}\cite{AI-Auth-Conf074}.
Others fail to develop a future roadmap \cite{AI-Auth-Conf133}\cite{Surv-Jour002}.
Most importantly, to the best of our knowledge, there is no relevant taxonomy in the literature.
Some existing surveys are briefly reviewed in the following.

The major conflict of an authentication system has been addressed: a high performance and a complex data pattern \cite{Surv-Conf003}. The researchers found that ANNs can recognize patterns accurately, but they have some limitations. Researchers reviewed the authentication using the finger vein pattern \cite{Surv-Conf001}. They found out that among other features like fingerprint, the finer vein pattern has the higher reliability.

Gupta et al\cite{Surv-Jour008} presented a description of ANN, and how it can help with authentication. They found out that methods based on ANN can provide higher accuracy and reliability than traditional password authentication.

A physical layer authentication method based on machine learning was presented for some devices that has limitation on storage and computing resources like IoT devices \cite{Surv-Jour001}. In their approach, the NN and SVM performed equally, and the OCC achieved a better accuracy.

Researchers compared 9 different deep learning models in different light conditions \cite{AI-Auth-Conf074}. They evaluate the models by the accuracy they gave, as well as the resources they need. Researchers compared the performance of the MLP and Cluster Analysis \cite{AI-Auth-Conf133}.

An approach that can authenticate food by using IR combined with multiple machine learning methods was provided in \cite{Surv-Conf005}. The researchers found out PLS performed an excellent accuracy among all tested models.

A survey was conducted by researchers on how the neural networks can help with the biometric authentication \cite{Surv-Jour002}. They came up with the future work such as use multiple biometrics for authentication at the same time to provide higher security. Researchers did comprehensive surveys on different type of deep learning models and nerual networks on how they can  contribute to the biometrics authentication \cite{Surv-Jour003}. They discussed the performance of each model for different biometrics like voice, faces \cite{Surv-Chap001}.

A mouse tracks authentication using 2D-CNN was presented \cite{Surv-Jour005}. This method use 2 Dimensional CNN to allow the transfer learning. By using SVM as a classifier, a high accuracy of authentication can be achieved. An authentication based on RNN has been produced to break the limit of AI based authentication \cite{Surv-Jour007}. The RNN methods gives us insights.

Researchers reviewed the biometrics authentication system by focusing on the advantages and disadvantages \cite{Surv-Jour009}. For future development, this paper pointed out that accuracy of authentication could be improved by combining it with the deep learning algorithms. Authentication and Authorization (AA) are key to CIA of IoT devices. The paper \cite{Surv-Jour010} proposed a taxonomy of authentication methods for IoT networks using machine learning. Researchers reviewed the behavioral biometrics based authentication schemes and gave the taxonomy of AI-assisted authentication methods \cite{Surv-Chap002}. In the paper, they focused on classification and clustering, and gave a road map for future development.

A scheme for fingerprint authentication was proposed by \cite{Surv-Conf002}. It combines the CNN and Siamese Neural Network to obtain the similarities between two images. The researchers found out that CNN and SNN highly improved the fingerprint authentication from traditional stage. Researchers reviewed four different schemes for physical layer authentication on 5G based IoT devices \cite{Surv-Conf004}. They provided the state-of-the-art and the future work such as combining the deep learning techniques to improve accuracy.

Elahee et al\cite{Surv-Conf006} summarized and collected the existing finger vein authentication methods based on deep learning. This paper compared the schemes in detail and gives the future research direction. Traditional mouse dynamic authentication schemes use simple machine learning tachniques to extract features. In \cite{AI-Auth-Jour047}, researchers proposed a new scheme for mouse dynamic authentication that extracting features from dataset using deep learning such as 2D-CNN. This module allow the transfer learning and achieved effectiveness on learning from a small dataset.

\begin{table}[]
    \caption{\textcolor{black}{Existing Approaches Comparison}}
    \label{tab:summary}
    \centering
    \begin{tabular}{ |m{1.3cm}|m{1.3cm}|m{1.3cm}|m{1.3cm}|m{1.3cm}|} 
\hline
 \textbf{Publicat-ion Year} & \textbf{Covers All AI-based Methods} & \textbf{Covers All Application Areas} & \textbf{Presents A Future Roadmap} & \textbf{Presents A Taxonomy} \\ 
 \hline
 1999\cite{AI-Auth-Conf133}  & No & No & No & No \\ 
 \hline
 2009\cite{Surv-Chap001} & No & Yes & Yes & Yes \\ 
 \hline
 2010\cite{Surv-Conf001} & No & No & No & No \\ 
 \hline
 2011\cite{Surv-Jour002} & No & No & No & No \\ 
 \hline
 2013\cite{Surv-Jour008} & No & No & Yes & No \\ 
 \hline
 2015\cite{Surv-Conf003} & No & No & Yes & No \\ 
 \hline
 2018\cite{AI-Auth-Conf074}  & No & No & Yes & No \\ 
 \hline
 2018\cite{Surv-Jour003} & No & Yes & Yes & Yes \\ 
 \hline
 2018\cite{Surv-Chap002} & Yes & Yes & No & Yes \\ 
 \hline
 2019\cite{Surv-Conf005} & Yes & No & No & Yes \\ 
 \hline
 2019\cite{Surv-Jour005} & No & No & Yes & No \\ 
 \hline
 2020\cite{Surv-Conf004} & No & No & Yes & No \\ 
 \hline
 2020\cite{Surv-Conf006} & Yes & No & Yes & No \\ 
 \hline
 2020\cite{AI-Auth-Jour047}  & No & No & Yes & No \\ 
 \hline
 2021\cite{Surv-Jour001} & No & No & No & No \\ 
 \hline
 2021\cite{Surv-Jour007} & Yes & Yes & No & Yes \\ 
 \hline
 2021\cite{Surv-Jour009} & Yes & Yes & No & Yes \\ 
 \hline
 2021\cite{Surv-Jour010} & Yes & Yes & Yes & Yes \\ 
 \hline
 2021\cite{Surv-Conf002} & No & No & No & No \\ 
 \hline
\end{tabular}
\end{table}

The summary of the reviewed papers are shown in \textcolor{black}{Table \ref{tab:summary}}. As seen in the table, we collected and compared the approaches that proposed from 1999 to present. Just few of them covered all the AI-based authentication methods. Only \cite{Surv-Jour010} covered all four compared items.

\section{State-of-the-Art}\label{State}

\subsection{Applications}

\subsubsection{Environments}

\paragraph{Mobile Devices}\mbox{}

Authentication in mobile devices has been growing rapidly, especially considering the fact that these small devices have many embedded sensors.The researchers use Cross-View security authentication using View-Invariant to overcome the previous Cross-View problems with previous authentication methods in \cite{AI-Auth-Conf058,a2}. 
Active frequent authentication is used to reduce the risk of someone getting access to an already unlocked mobile device \cite{AI-Auth-Conf090}.
 Researchers proposed using facial attributes \cite{AI-Auth-Conf161} such as hair colour, gender et cetera to authenticate users, this approach reduces the complexity of the network.

\paragraph{IoT Environments}\mbox{}

Internet of Things has been rapidly growing \cite{a9, a10}. The number of smart devices has grown exponentially. The authentication of IoT devices is closely related to the security of users.

When it comes to on-body IoT devices like smartwatches, it is hard to authenticate under diverse user motions and complex IoT network environments. A new approach, with the help of the adversarial multi-layer neural network, wireless physical layer and the upper layer protocols were used to authenticate the wearable devices.\cite{AI-Auth-Jour001} Based on RSS, the experiment resulted with authentication accuracy of 90.6\%. The proposed method can significantly reduce attack on wearable IoT devices.

As for privacy concerns, a recent research on the authentication of IoT devices involved blockchain technology\cite{AI-Auth-Jour014}. A trustworthy blockchain was built using this approach to protect privacy, due to the elevating risk of attacks on IoT devices. Transfer Learning has been used to improve the authentication scheme. Their result tells us that this method achieved very accurate IoT authentication. Simultaneously, it provides a very fast processing speed with low latency.

In the IoT world, low-power devices like sensors rely on some high-power devices to emit signals to achieve communications. Sometimes it is hard to authenticate the low-power devices with a lot of strong, powerful signals around\cite{AI-Auth-Conf044}. A solution has been given by Das et al\cite{AI-Auth-Conf044}. They created a model with deep learning to classify the I/Q signal streams. So that they can correctly identify the low-power devices and start the authentication process.

When IoT devices are manipulated by some unauthorized person, some severe problems can be caused. To avoid tragedy, we need to detect and thwart suspicious control. According to recent researches, we can try to detect misuse by using biometrics \cite{AI-Auth-Conf099}. Karimibiuki et al came up with an autonomous IoT system behaviour self-check method based on KNN, SVM and LR\cite{AI-Auth-Conf092}. They use machine learning to classify the human operation and device autonomous behaviour. When some behaviours are out-of-path, it can block them right away to ensure safety.

\paragraph{Controller Area Networks (CANs)}

The Controller Area Networks are used in most vehicles to share data between different parts of the vehicle. The authors of \cite{AI-Auth-Jour021} have used an authentication scheme that analyzes the intervals between messages and voltage to detect an attack on CANs. The authors of \cite{AI-Auth-Conf060} have also used voltage signals as an authentication method, a similar approach to\cite{AI-Auth-Jour021}.

\paragraph{Sensor Networks}

In terms of sensor networks, a research that has been doe by Xiao et al shows a physical layer authentication method for Underwater Sensor Networks by using machine learning\cite{AI-Auth-Jour024}. In their approach, they set up a reinforcement learning(RL) model to deal with the sensor spoofing attack. They built up a formula to calculate a Q-value each time sensors update information to the data center. If that value falls into an unacceptable range, there will be a spoofing alarm. After using the RL method to calculate the Q-value with a formula, they came up with the DRL method which involves the neural network. The Convolutional Neural Network(CNN) has been used to accelerate the calculating process. By replacing the static formula with CNN, the accuracy of alarm has been significantly increased.

\paragraph{Vehicular Ad hoc NETworks (VANETs)}

Vehicular Ad hoc NETworks are vehicles connected to a wireless network \cite{a11,a12}. A new authentication method is proposed for VANETs \cite{AI-Auth-Jour042}, this method uses intervals to detect packet arrival that could be sent for spoofing.

\paragraph{Medical systems}\mbox{}

As to medical systems, there are some limitations on the selection of authentication features. Usually, users of medical systems are going through different physical or mental illnesses. For instance, some users may not be able to memorize, some users may be unable to see, speak or move their bodies.

Some recent discoveries show some authentication methods based on the biometrics features such as iris patterns. Researchers use Hierarchical CNN to assist iris recognition authentication \cite{AI-Auth-Conf160}. 
A new iris recognition authentication method was introduced \cite{AI-Auth-Conf128}. The scheme uses YOLO v2 deep learning to combine the Iris and Sclera patterns. The algorithm makes decisions by considering partial Iris and Sclera, which increased the authentication accuracy.

However, the authentication based on iris recognition can not work for some situations, such as for patients with eye illnesses who are unable to open their eyes. To solve this problem, the direction of research should be searching for some features that apply to almost everyone.

An authentication method using electrocardiogram(ECG)-based biometric information has been proposed in \cite{AI-Auth-Jour033} 
.  Researchers presented four new schemes for various ECG occasions. In addition, they developed a MATLAB toolbox for their method for later use. Researchers proposed an ECG-based authentication for wearable devices to ensure healthcare data security \cite{AI-Auth-Jour033}. This method uses ANN as the second phase of authentication. The speed of the authentication process is significantly surged up with the help of deep learning. Introducing by implementation, the FRR and FAR of their approach are both below 10\%. An authentication method for inconvenient people like patients with Dementia was presented by Farid and Ahamed
. Their approach allows those patients to reach their healthcare data by using ECG and PPG-based biometrics. With the help of Recurrent Neural Network, their method reached a high accuracy.

\paragraph{Virtual Reality Environments}\mbox{}

Virtual Reality is currently used in many different ways, delivery of education, entertainment et cetera. The authors of \cite{AI-Auth-Conf151} have introduced a new authentication method that uses Siamese Neural Networks to authenticate users. This method uses physical features while using the technology. Some of the many ways used to authenticate users is the way a person throws a ball using the VR controls.

\paragraph{Educational Systems}\mbox{}

Virtual Reality gives us more ideas on education. For those students who major in 3D modelization, Physiology and Medicine, VR gives them chance to view the inner structures of a machinery component, the human body and the Molecular spatial structure. But at the same time, VR collects personal biometric information like motions, which should be kept confidential. Therefore, it is important to figure a way to authenticate the person. There are some approaches that verify a person by password and 2FA, but they are too easy to be misused or stolen. Thus, Miller et al came up with a method that uses Siamese Neural Networks to perform a  cross-system authentication\cite{AI-Auth-Conf151}. According to their research, their algorithm can distinguish the biometrics of a person from different VR platforms. And their work achieved an accuracy of 98.53\%.

\subsubsection{Networking Layers}

\paragraph{Physical Layer}

The authors of\cite{AI-Auth-Jour002} use Transfer Learning based PHY-layer Authentication that authenticates online users faster.

The authors of \cite{AI-Auth-Jour012} have argued that threshold in hypothesis is not always reliable. Therefore, they proposed a new physical-layer to authenticate dynamic-networks since old methods such as comparing radio-channel frequencies that improves reliability and robustness \cite{AI-Auth-Jour023}.

Physical authentication for mobile scenarios has been difficult with existing PHY-AUC. The authors of \cite{AI-Auth-Jour017} proposed a threshold-free PHY-AUC method that replaces the old methods and imposes a new method using classification based machine learning. But \cite{AI-Auth-Conf095} argued that threshold optimization could provide authentication by recording the performance of each potential attribute. 

Users can now be authenticated by utilizing the changes in sub-carrier amplitude information that helps identify users using passive Wi-Fi sensors\cite{AI-Auth-Jour019}. A problem with authentication many threats can not be detected using cryptographic methods due to their limited resources\cite{AI-Auth-Conf101} 
. The fix for this is Neural Network \cite{AI-Auth-Jour039}\cite{AI-Auth-Jour051} is used as an intelligent authentication process to improve authentication and reduce risks. Machine learning was also used as a physical layer authentication \cite{AI-Auth-Conf028}\cite{AI-Auth-Conf052} \cite{AI-Auth-Conf091}.

\paragraph{Other Layers}

Only a few researchers have studied AI-based authentication in other layers.

There is very rare researches have been doe to implement the AI-based authentication in other layers. Here are some examples in the Network layer and Application Layer.

\begin{itemize}
    \item \textbf{Network Layer:}

In the Network Layer, secure network transmission is constrained for many IoT devices due to the limited resources\cite{a13, AI-Auth-Jour004}. Physically Unclonable Functions solved the cryptography problem during the network transmission, but it is vulnerable to Machine learning and man-in-the-middle attack. This work proposed an anonymous authentication protocol on the scalable level which can reduce the risk of being attacked.\cite{AI-Auth-Jour004} Especially, their approach was designed for the Internet of Medical Things (IoMT) to insure the safety of medical data.

    \item \textbf{Application Layer:}

In terms of the Application Layer, biometrics like facial features and behaviours are widely used in AI-based authentication. Conventional facial recognition used simple and static image processing algorithm. The accuracy and processing speed are the barriers. But when it comes to machine learning, these core problems have been solved. Researchers came up with a foot image recognition system to authenticate a person on some specific occasion like in hospitals \cite{AI-Auth-Conf067}. Their approach used SVM, and deep learning models and reached a high accuracy of 98\%.

Smartphones become one of the necessities of people's life. A variety of sensors on the phone are recording people's behaviour almost every time. Hence, researchers are recently focusing on how to authenticate a person through his behaviour on smartphones. In the research of  Irvan et al, a proposed model of behaviour authentication can learn from a large number of possible patterns \cite{AI-Auth-Conf050}. They demonstrated a possible way to authenticate users by the way they touch and swipe their smartphones.
\end{itemize}

\subsubsection{Authentication Methods}

\paragraph{Biometric Authentication}\mbox{}

\textcolor{black}{Biometric authentication refers to the authentication methods that use human beings' biological characteristics and behaviour. Biometric authentication has always been involved as a hidden layer of authentication, as well as the second layer to strengthen security.}

Many researchers have focused on the applications of AI in biometric authentication \cite{AI-Auth-Conf169}. Among different types of biometric authentication using AI-based methods, we can mention the following ones.

Biometric Authentication is divided into two types, physical authentication and the second type is behavioral authentication. Both types are used to authenticate a person with unique characteristics. Physical Biometric Authentication use distinctive characteristics such as faces, fingerprints, DNA,etc. Whereas behavioral Biometric Authentication uses voices, way of typing, etc. Collected data is then transformed into code that is understandable by the AI to process.

\begin{itemize}

	\item \textbf{Face Authentication:}
	
	\textcolor{black}{Face authentication is a series of methods that can authenticate people by analyzing their facial features. Some of the facial features used are, shape of eyes, the distance between nose and lips. These features are easily captured and can be very efficient for machine learning. Thus, authentication with facial recognition is very popular.}
	
	Face authentication is the process where images and videos are captured of human faces to identify a person from a specific database\cite{AI-Auth-Conf015}. Face authentication method focuses on face prints which are unique for every person. Some of these face prints are the size of the eyes, the distance between the eyes, size of the nose, head shape,etc. Usually face recognition is used on 2D images and videos which impose some problems but a study shows how using a DSLR camera, DLT and neural network algorithm can make it easy to detect 3D faces\cite{AI-Auth-Conf108} . This method prevents access using a photo of a person which is 2D and only detects 3D live faces\cite{AI-Auth-Conf121}. Another issue is detecting people of colour \cite{AI-Auth-Conf116}. Deep Reinforcement was used for facial recognition\cite{AI-Auth-Conf070}. It was also used for a self adapting Deep learning algorithm\cite{AI-Auth-Conf065}. The authors of \cite{AI-Auth-Conf009}  have added an addition function using Deep learning which is auto-sharing images using facial recognition of people in a picture\cite{AI-Auth-Conf014}.

	\item \textbf{Typing Biometric Authentication:}
	
	\textcolor{black}{Typing biometric authentication or keystroke biometric authentication is a newly emerged method that leverages people's typing features such as typing speed, the time duration between hitting two keys.}
	
	Typing Biometric Authentication, is a type of authentication that uses physical characteristics and calculates them to authenticate a user. Typing biometrics are used to authenticate users nowadays. A paper suggests using artificial neural networks and k-nearest neighbor algorithm to authenticate users through typing. This is doe by calculating the time intervals taken between each key stroke while entering a password, This method is combined with using a password for authentication which makes it hard for other users to get access since they do not share the same typing patterns	\cite{AI-Auth-Conf127}.

	\item \textbf{Behavioral Biometric Authentication:}
	
	\textcolor{black}{Behaviour biometric authentication refers to the methods based on the pattern that was shown when a person interacts with devices like a smartphone. It identifies people by behaviours like swiping vectors, keystroke dynamics, and mouse moving tracks.}
	
	Behavioral biometric authentication depends on how a user behaves. From key strokes to time intervals between every click, touchscreen behaviour and device movement for portable devices are all types of behavioural biometrics that could be analyzed to provide an extra stronger layer of security	\cite{AI-Auth-Jour003}. These behaviours are fed into a machine learning model and analyzed to authenticate users better than physical characteristics that could be bypassed. A recent research shows how an embedded PC authentication could save users authentication time and better authenticates by authenticated the user behind the scenes\cite{AI-Auth-Conf104}.

	\item \textbf{Biometric Signature Authentication:}
	
	\textcolor{black}{Signature authentication refers to identifying people by their handwritten signatures. People believe that handwritten signatures are unique for each person and hard to be perfectly mimicked. Therefore, signatures have been used to identify people a long time ago. With the development of AI, some neural networks are used for identifying peoples' signatures to achieve high accuracy and advanced security.}
	Biometric Signature Authentication uses signatures to authenticate users and avoid relying only on a password that could be easily stolen nowadays. A research paper suggests the use of recurrent neural network to authenticate users using their signatures\cite{AI-Auth-Conf042}.

	\item \textbf{Fingerprint Authentication:}

	\textcolor{black}{Fingerprint authentication is a technology to identify people by scanning their fingerprints. Researchers believe that it is highly unlikely that two people share one fingerprint pattern. In addition, fingerprints are one of the easiest biometrics to get from a person. Thus, fingerprint authentication has been leveraged for decades.}
	
	Fingerprint Authentication uses human's fingerprints to authenticate users by scanning the fingerprint \cite{AI-Auth-Jour018}. It can be validated with the saved fingerprint in the device	 	\cite{AI-Auth-Conf024}.

	\item \textbf{Finger Knuckle Authentication:}
	
		\textcolor{black}{Finger Knuckle Print (FKP) authentication is a way that authenticates a person by finger knuckle patterns. This method usually detects the tiny details of finger knuckles with the help of radiation. }
	
	Finger Knuckle Authentication is a new and promising way to authenticate users. Finger knuckles are scanned for user authentication, finger knuckles are used because the bending of the knuckle is very unique which provides a high level of security and reduces the chances of false-positives. A paper shows how using the left index knuckle print and the SCG algorithm can provide authentication with 96.03\% accuracy \cite{AI-Auth-Conf153}.

	\item \textbf{Body Odor Authentication:}
	
	\textcolor{black}{There are many unchangeable features in the human body. DNA, fingerprints, iris patterns are the most well-known. However, body odor is also classified as an unchangeable human body feature. Since body odor is unique for every person, authentication can be done by examining the body odor of users.}
	
	Body Odor Authentication uses a gas sensor to authenticate users based on their body odor and machine learning \cite{AI-Auth-Conf072}. In a recent research paper they used three different machine learning algorithms to achieve this, K-means, Principal Component Analysis, and Neural Network were the algorithms used \cite{AI-Auth-Conf076}.

	\item \textbf{Voiceprint Authentication:}
	
	\textcolor{black}{People speak a lot during the day. The voice pattern of a person normally does not drastically change during a long time span. Voiceprint authentication, also known as speaker recognition, is a way to validate the user's claimed identity by the features extracted from their voice. }
	
	Voiceprint authentication uses human voices to authenticate users, this has been used recently in banks, and governments. Voice recognition algorithms are used when calling to perform banking transactions, or any other request personal information. In most cases this is not the only way of authenticating the user, customers are asked questions to answer to and then their voice is processed and authenticated with previous calls to confirm identity. This authentication is done through collecting the voice mel-frequency Cepstral coefficients and tested and authenticated using a machine learning model\cite{AI-Auth-Conf100}.

	\item \textbf{Multimodal Biometric Authentication:}
	
	\textcolor{black}{It is considered a multimodal authentication system when two or more biometric types are available to choose from when the system identifies a person. Authentication schemes that involve facial recognition and voice recognition, a smartphone that accepts fingerprint and behavioural biometrics are falling under the multimodal.}
	
	Multimodal authentication uses more than one biometric. This comes in handy since one biometric authentication is not the most secure \cite{AI-Auth-Conf113}, therefore combining two biometric authentication \cite{AI-Auth-Conf020}. A 2018 research paper suggested using convolutional neural network where biometric information is infused together from different resources \cite{AI-Auth-Jour055}.

\end{itemize}

\paragraph{Password Authentication}\mbox{}

\textcolor{black}{Passwords have been the primary authentication for many years. Users should choose a passcode or passphrase only known by them to authenticate their credentials.}
Controversial authentication methods based on passwords are out-of-date. Traditionally, when users register, the server will record the credentials like usernames and password fingerprints. When it comes to logging in, the server will compare the received password and the record in the database. This method is not the most secure as the server stores all users' credentials. Using this method could result in data leakage, or misue.

A new password authentication scheme has been created by Li et al.\cite{AI-Auth-Jour048} Their approach explained a way that multiple servers can authenticate an entity with the help of ANN. According to their research, the servers only store a weight of a password. Multiple servers will store their own part of the weight, then by using a pattern classification system, the servers can make a decision on if a user passes the authentication.

\textcolor{black}{Hopfield Neural networks (HNN) can help with password authentications. Hopfield neural networks are special types of recurrent neural networks introduced by Little in 1974 and popularised by John Hopfield in 1982. It provides a model to understand human memory. } There is a research has doe by Wang et al\cite{AI-Auth-Jour044} shows how Hopfield Neural Network can help on this. Their works achieved higher accuracy and less response time delay than other existing neural network approaches. Similarly, A Hopfield Neural Network with fuzzy logic authentication has been come up in 2018.\cite{AI-Auth-Conf105} Researchers added the fuzzy logic feature to improve the accuracy of previous work.

However, after Li et al\cite{AI-Auth-Jour048} coming up with their method, in 2001, Wei-Chi Ku found out that there are some vulnerabilities that are not repairable in that scheme \cite{AI-Auth-Jour046}. For Li's scheme, the offline guessing attack may be a threat. Besides, there is no countermeasure that can deal with the insider attack. In terms of password attacking, offensive security researchers have proposed an attacking method\cite{AI-Auth-Jour025} that can more efficiently steal users' credentials.

\paragraph{CAPTCHA Authentication}\mbox{}

CAPTCHA authentication is used to distinguish between humans and machines. This is doe by using a neural network machine learning model. A study used CAPTCHA to verify smart-TV users by running a machine learning model that distinguishes between real human images and downloaded images to limit attacks on smart-TV and other IOT devices\cite{AI-Auth-Conf022}.

\paragraph{Mouse Dynamics Authentication}\mbox{}

Mouse dynamics authentication is a series method to authenticate an individual by analyzing the mouse movement and dynamics. Some recent works have shown the authentication by mouse movements\cite{AI-Auth-Conf084} or by mouse dynamics or by both. A mature dataset has been presented by Antal et al\cite{AI-Auth-Conf034} for the public to train the models.

To achieve authentication by mouse features,\cite{AI-Auth-Conf087-1} proposed a basic machine learning model by Weighted Least Square Regression(WLSR) and Learning Vector Quantization(LVQ). \cite{AI-Auth-Conf053} used RNN combined with users' eye detection. \cite{AI-Auth-Conf062} introduced a semi-supervised learning model which achieved  0.78 AUC and 0.26 EER as a result.

\paragraph{Keystroke Authentication}\mbox{}

\textcolor{black}{Keystroke and keystroke dynamic has been leveraged into authentication. When people type, the typing speed, time duration of hitting and releasing a key, the strength that they hit a key can fully represent their personalities. Keystroke authentication refers to the approach that authenticates people by recording and analyzing their keystroke patterns. }

Keystroke authentication is the authentication that depends on the timing of each key press and release. A method presented in \cite{AI-Auth-Conf106} Shows how identity is authenticated when inputting the user password and denying access once it detects unusual keystroke pattern. Neural network was used \cite{AI-Auth-Conf122} to authenticate keystrokes by investigating the performance of keystroke dynamics and using multilayer perceptron\cite{AI-Auth-Conf139}.

\paragraph{Hand Motion Authentication}\mbox{}

Hand motion is easy to be detected. So it has been used for authenticating a person in multiple types of systems. According to the research that has been conducted by Shioji et al\cite{AI-Auth-Conf103} in 2018, it is easy to achieve authentication by either hand motion or EMG. However, it is hard to combine them together. In their work, they use multiple inputs and outputs CNN with 3×3 convolution to combine those two methods. And they reached an accuracy of 95\% on average.

\paragraph{Hand Gesture Authentication}\mbox{}

The use of hand gesture authentication is very limited to user-friendliness and how accurate it is as it required a specific hand gesture to be doe to authenticate a user. In \cite{AI-Auth-Conf148} researches have implemented a new way to authenticate users which is by performing an improvised hand gesture that is then utilized by a machine learning model to authenticate users.

\paragraph{Ankle Motion Authentication}\mbox{}

Gadaleta et al\cite{AI-Auth-Conf145} proposed a way to authentication a person by an ankle-worn sensor. Researchers use CNN as a feature extractor to analyze the ankle motion data and feed them to the SVM classifier. According to their approach, they can very accurately authenticate a person with false rates under 0.01.

\paragraph{Gait Activity Authentication}\mbox{}

Gait activity authentication is doe by reading the locomotion of humans, which is the movement of a person. A research paper have come up with a new way to authenticate users by using gait activity using smartphone sensors and LSTM neural networks which analyzes human activity and extracts it \cite{AI-Auth-Conf159}.

\paragraph{Handwriting Authentication}\mbox{}

Handwriting is a widely used identification method since ancient times. Traditionally, handwriting should be recognized by human beings. However, recent research shows that handwriting can be verified by machine learning algorithms. Hayashi et al proposed a model that combined both the handwriting and writing dynamic data to authenticate an individual.\cite{AI-Auth-Conf110} When a person gives his handwriting, the machine uses Siamese Neural Network for handwriting tracking recognition. At the same time, it will analyze the dynamic of writing by Multilayer Perceptron. In conclusion, they found out that the authentication accuracy of their approach is higher than using either tracking recognition or dynamic analysis.

\paragraph{Handwritten Signature Authentication}\mbox{}

Handwritten signature authentication is a very ancient way of authentication, physical handwritten signatures are used in bank up until this day \cite{AI-Auth-Conf138}. Artificial Neural Network was used to verify handwriteen signatures \cite{AI-Auth-Conf144}. the authors of \cite{AI-Auth-Conf114} used  Canny edge detector which is an image filter that detected and authenticated users offline. Another method used speed of writing, order of writing and skilfulness to analyze and authenticate a signature \cite{AI-Auth-Conf146}.

\paragraph{Activity Authentication}\mbox{}

Activity authentication is a way that allows machines to verify a person by analyzing the daily motion data captured by a  group of sensors. With the number of smartphones surging up, it is a lot easier to collect daily motion data than ever before. Ahmadi et al\cite{AI-Auth-Conf055} proposed a method to authenticate a person by classifying the motion records such as walking, standing, sitting, going upstairs and downstairs.

\paragraph{Multifactor Authentication}\mbox{}

Multifactor authentication refers to the science and technology of authenticating an object using more than one authentication factors.

Generally, we have three types of factors to help with authentication. For instance, we can be verified by the passwords we know, the ID card we have, or the features of our own, such as facial characteristics. When we combine at least two of those factors, the authentication is called two or multi-factor authentication.

The method in \cite{AI-Auth-Conf137} uses CogRAM Weightless Neural Network to authenticate individuals by the passwords and the dynamics of keystrokes when they input their passwords. The scheme proposed in \cite{AI-Auth-Conf141} is a signature-based verification bound with mouse movement authentication.

\paragraph{Hybrid Authentication}\mbox{}

Hybrid authentication refers to the science and technology of authenticating an object using more than one authentication method.

Traditional biometric authentication such as face, iris recognition uses machine learning to identify a person. For further security, the device itself need to be authenticated at the same time. Hence, Banerjee and Ross \cite{AI-Auth-Conf064} proposed a scheme that joint the user identification and device authentication together during a biometric authentication process. Their approach uses an embedded network which contains 2D-Convolution and Linear modules to reach a high accuracy.

\subsubsection{Objects to Be Authenticated Objects}

In this section, we reviewed authentication methods used to identify different objects. Especially in IoT world, to authenticate and trust a device is of importance.

AI-based methods have been used to authenticate a variety of objects, some of which are discussed below.

\paragraph{Person Authentication}\mbox{}

The literature comes with several research works focusing on AI-assisted person authentication \cite{AI-Auth-Conf067}. Depending of the role of the person to be authenticated, there are different types of person authentication \cite{AI-Auth-Conf016}. A couple of well-studied types of AI-based person authentication are briefly discussed in the following.

\begin{itemize}
	\item \textbf{User Authentication:}

	User Authentication has been a very sensitive subject lately as there is a lot of research happening around it. The authors of 	\cite{AI-Auth-Jour011}  used mouse dynamics to authenticate users. To prevent mouse data from an attack, mouse positions were randomized to make it harder for an attacker to detect the actual mouse pattern \cite{AI-Auth-Jour047}.Some approaches had to be build on a budget considering processing power \cite{AI-Auth-Conf083}. Continuous authentication was used\cite{AI-Auth-Conf111}. Keystroke authentication \cite{AI-Auth-Conf085} is proposed using robust recurrent confidence model and authentication using hand motion and movement gestures. The authors of 	\cite{AI-Auth-Jour026} also rely on keystrokes to authenticate users, the force and touch duration are extracted using the piezoelectric force touch panel.

	\item \textbf{Driver Authentication}

\end{itemize}

\paragraph{Person Authentication}\mbox{}

\textcolor{black}{AI-assisted device authentication has been of interest to the research community \cite{AI-Auth-Conf048} . Recent search shows that different kinds of devices can be authenticated using AI-based methods. Some relevant works are categorized and briefly studied below.}

\begin{itemize}
	\item \textbf{Wireless Device Authentication:}
	
In the IoT world, traditional device authentication based on complex cryptography is hard to implement due to the limited computing and processing of the devices. The authors of  \cite{AI-Auth-Jour030} provided a method based on transfer learning to identify a device using fingerprint information. The proposed scheme can detect the physical attack towards the fingerprint while traditional authentication methods can not do that. There is an approach \cite{AI-Auth-Jour038} to help the authentication of IoT wireless devices by using In-situ machine learning. The method they introduced uses radio-frequency and Physical unclonable functions to uniquely authenticate each die. Both of these approaches are able to process authentication very fast with the consumption of a low resource.
	
	\item \textbf{Transmitter and Transceiver Authentication:}
	
	A transmitters authentication method was provided in \cite{AI-Auth-Jour050}. Researchers compared the functionality of Recurrent Neural Network (RNN) with Long-short Term Memory (LSTM) and RNN with Gated Recurrent Unit (GRU) to authenticate the transmitters in a mobile situation. In terms of transceiver authentication, \cite{AI-Auth-Conf093} proposed a scheme based on physical layer authentication. The model can extract the specific device behavior called CFO to train the classifier to identify whether a transmitter is legitimate or not. As to validate the transceiver, the scheme uses SDR to record the relevant measurements. The SVM gives the best accuracy and performance among other adopted classifiers.
	
	\item \textbf{Emitter Authentication:}

	As the variety of IoT devices are emerging in our life, they begin to collect and transmit some sensitive information. Hence, it is important to authenticate and identify an emitter.
	
McGinthy et al\cite{AI-Auth-Jour053} found that Neural Networks are able to help identify a transmitter according to the waveform characteristics. They have doe a groundwork
	for emitter authentication by using NNs.
	
	\item \textbf{IC Authentication:}

	Integrated Circuits(IC) need to be authenticated as well \cite{a14}. Based on the type of Analog-to-Digital Converters (ADCs), \cite{AI-Auth-Conf002} proposed a 3-layer ANN to identify and authenticate different ICs.
	
	\item \textbf{Sensor Authentication:}

Murphy et al\cite{AI-Auth-Conf032} proposed a sensor authentication by using Hidden Markov Models(HMMs). They found out their model has a remarkable stability.
	
	\item \textbf{Handheld Device Authentication:}

Conventional iris detection authentication uses basic recognition methods with intermediate accuracy.  \cite{AI-Auth-Conf033}  explained the importance of accurate segmentation in  the process and proposed a new iris authentication scheme with deep learning algorithm to help accurately segment and identify iris patterns.
	
	\item \textbf{Smart Phone Authentication:}

Existing smartphone authentication methods are not satisfactory enough both in security and in the amount of data they collected. The new methods proposed in \cite{AI-Auth-Conf049} equipped with deep learning algorithm to learn from the dataset which only contains the accelerometer data. According to their implementation, the proposed model can reach an ERR as low as 2.2\%.
	
	\item \textbf{Mouse authentication:}

Traditional mouse authentication methods uses general machine learning to obtain the features of the track patterns. The scheme introduced by \cite{AI-Auth-Conf056} uses a CNN model to extract features from patterns, and uses a 2D-CNN to predict. This approach provides a higher accuracy and a good performance to deal with curve crossing.
	
	\item \textbf{IoT Device authentication:}

    A taxonomy was provided for IoT devices authentication using machine learning \cite{AI-Auth-Conf071}. IoT devices are emerging and ubiquitous. The security issues need to be addressed along with its development. However, there are some limitations of IoT devices, such as limited computing and processing abilities. Their work gave us a taxonomy of the security approaches towards IoT devices and showed future directions.
	
	Bari et al\cite{AI-Auth-Conf063} provided a method to authenticate the IoT devices by Radio Frequency (RF) with Incremental clustering. According to their experiment using 30 Xbee modules, they got high accuracy and low latency result. Their approach provides extra security and does not need additional protection when combining their algorithm with existing schemes.
	
	\item \textbf{Speaker Authentication:}

	When different people speak, their lips' movements will be unique. Shi et al\cite{AI-Auth-Conf098} provided an authentication method for speakers by using lip features. In their approach, they recorded the lips' movements as video clips and extract features from them. An SVM model was trained to classify the samples and make decisions. Their implementation result is better than those produced by traditional methods. Two years later, in 2018, Liao et al\cite{AI-Auth-Conf135} presented a 3D-CNN-based authentication model for speakers, which focused on the lips' motion as well. This new work adopted the 3D-CNN to learn features in a complex way. The experiment demonstrated an excellent accuracy of 99.18\%.
	
\end{itemize}

\paragraph{Signal Authentication}\mbox{}

Signal authentication is used to authenticate wireless signals sent to IOT devices.In \cite{AI-Auth-Jour037} they used a watermarking algorithm, to get stochastic features of the signal. After features are collected signals are authenticated to avoid any cyber attacks on IOT devices. Deep reinforcement learning algorithm was used and it reduced cyber attacks by up to 30\%.

\paragraph{Content Authentication}\mbox{}

AI-based content authentication has received attention from researchers \cite{AI-Auth-Jour043}.

Content authentication is a group of theories and techniques that ensure the integrity of content like digital photos, videos. And to determine if a media object was modified. It can also authenticate the identity of the contents' authors to protect copyright and ownership.

The broad term of AI-based content authentication covers different kinds of contents, some of which are studied in the following.

\begin{itemize}
	\item \textbf{Image Authentication:}
	
The method presented by \cite{AI-Auth-Conf019} uses extreme learning machine and Zernike moment to generate a robust digital watermark that can resist multiple types of attack \cite{AI-Auth-Conf115}. This approach can be used for copyright authorization.

	\item \textbf{Video Authentication:}
	
The research has been doe in \cite{AI-Auth-Conf054} is to authenticate a video by detecting the tampering attack using SVM to classify the frames. The scheme can detect the modification in both temporal and spatial fields. Besides, it gives an excellent authentication accuracy of 99.12\%.

	\item \textbf{Multimedia Content Authentication:}
	
The multimedia authentication method in \cite{AI-Auth-Conf005} breaks the limitation of out-of-distribution data that most existing works have. The proposed method focuses on the reliability and useability of the framework. Bayesian neural network (BNN) has been used for estimating and also giving an uncertainty range. Besides, this framework can detect unseen JPEG compression and resampling.

\end{itemize}

\paragraph{Message Authentication}\mbox{}

Message authentication is used to provide integrity and make sure text has not been altered. Physical layer based authentication \cite{AI-Auth-Conf059} is used with supervised learning. Channel based Security can be achieved, this is doe by processing user specific characteristics \cite{AI-Auth-Conf006}.

\paragraph{Authentication of Other Objects}\mbox{}

In addition ot the above objects, researchers have worked on the authentication of other objects such as bank notes \cite{AI-Auth-Conf021} and herbal leaves \cite{AI-Auth-Conf023} using AI-based methods.

\subsection{Attack and Defense}

In this subsection, we study research works proposing AI-based methods for attacking authentication systems or protecting them against attacks.

\subsubsection{Attack}

The authors of\cite{AI-Auth-Jour005}  use Authnet. Authnet is a framework that uses biometrics to authenticate users \cite{AI-Auth-Conf037}, and have shown very successful results using Authnet which regularizes mapping space and improves authentcation accuracy. Deep learning based attacks were used \cite{AI-Auth-Conf026} the researchers were able to train a machine learning model to identify facebook accounts by using a web crawler and using the information to get access. The authors of \cite{AI-Auth-Conf013} used a similar machine learning attacks using deep learning.

\subsubsection{Defense}

The attacks are emerging every second, and traditional protecting systems require human to response when an attack happens. With the help of machine learning, in some occasion, the system itself will be able to deal with attacks automatically \cite{a15}.

To protect IoT devices from being attacked, the Physically-Unclonable-Functions(PUFs) were implemented. But generating unique and random fingerprints when communicating with each other are vulnerable to machine learning-based attacks. Some protocols \cite{AI-Auth-Jour004} that allow anonymous authentication and challenge-response schemes can protect PUFs forge attacks from machine learning \cite{AI-Auth-Jour036}.

To avoid machine learning-based attacks, password-based authentication has been gradually eliminated in IoT networks. \cite{AI-Auth-Jour015} introduced a challenge-response scheme to prevent learning-based advanced attacks. It makes IoT devices safe when used in the public network.

DoS attack is a common vulnerability on the MAC layer. Servers have been and will continue to be the target of this kind of attack. In \cite{AI-Auth-Conf068}, researchers proposed a machine learning-based IPS to monitor the traffic and alert when an attack has been performed. By combining with machine learning, the system can identify malicious traffic and block them automatically. For all devices working under the 802.11 protocol.  Existing solutions are not able to accurately locate the attackers. The approach in \cite{AI-Auth-Conf077} is a detection system based on machine learning. It allows us to detect the DoS attack towards the Wi-Fi network and learn from the clients' behavior to decide who is the possible attacker. For both clone and Sybil attacks, \cite{AI-Auth-Jour022} presented a method to detect by channel-based machine learning. Combining with edge devices, their method can achieve an accuracy of 84\% without human labelling.

\subsection{Enabling Sciences and Technologies}

\subsubsection{Statistics and Probabilistics}

Researchers showed that the use of multimodal biometric authentication and feature vectors results in a very high practical value \cite{AI-Auth-Conf102}. In another research \cite{AI-Auth-Conf047}  one-class nearest neighbor (OCNN) was proven to provide better security performance over statistical methods.

\subsubsection{Information Theory}

Shannon's information theory can help us build an information fusion model\cite{AI-Auth-Conf134}. This approach uses metrics to compute the identity attributes and performs authentication by information fusion.

\subsubsection{Smart Phone Technology}

AUToSen was proposed by researchers\cite{AI-Auth-Jour034} to provide continuous smart phone authentication that is based on deep-learning. AutoSen uses phone sensors such as accelerometer, gyroscope, and magnetometer to authenticate users and has shown very promising results for authentication.The authors of \cite{AI-Auth-Conf050} have come up with a new method of authentication by implementing a customized learning classifier systems, which can authenticate users based on their usage patterns of specific mobile applications.

\subsubsection{Embedded Processing}

There are a lot of sensors for authentication embedded in wearable devices. For patients, there is some physical or mental limitation to be authenticated using traditional methods. \cite{AI-Auth-Conf154} introduced a scheme that can authenticate a person by ECG pattern. The sensor was embedded into a wearable device.

\subsubsection{Sensor Technology}

Sensor Technology is used by the researches of \cite{a23, AI-Auth-Conf039} 
 that use behavioral characteristics, which uses built in smartphone sensors, and context-based authentication models that provides authentication with 98.1\% accuracy. Deep learning was used in addition to a passive infrared sensor that is used to authenticate users based on their captured chest motion \cite{AI-Auth-Conf061}.

\subsubsection{Data Compression}

Data compression can help reduce the number of features. In \cite{AI-Auth-Conf081}, researchers use three different compression methods to shrink the data features so that the machine learning algorithms can perform efficiently.

\subsubsection{Blockchain}

Researchers are now explioting Blockchain to authenticate users and provide and extra layer of security \cite{a16, a17}. Researchers proposed the use of an authentication mechanism empowered by Blockchain coined ATLB that is based on Transfer learning, which provides accurate authentcations for Industrial Internet of Things (IIoT) applicaton and provides high throughput and low latency\cite{AI-Auth-Jour014}.

\subsubsection{Neural Cryptography}

Neural cryptography can be used in key management such as key exchange processes \cite{AI-Auth-Conf004} and key management protocols \cite{AI-Auth-Conf164}. The method \cite{AI-Auth-Conf004} proposed is based on secret boundaries, while researchers involved Public Blockchain \cite{AI-Auth-Conf126} in the randomization process.

There are some other shreds of evidence for neural cryptography implementation. It was deployed on WiMAX authentication \cite{AI-Auth-Conf117}. In some practical authentication scenarios, such as multimedia data authentication using Cyclic Elliptic Curve \cite{AI-Auth-Conf147}. Besides, it is popular in the IoT fields. For instance, it helps IoT devices authentication, encryption and key management \cite{AI-Auth-Jour045}. Also in security-enhancing for IoT wireless nodes\cite{AI-Auth-Conf082}. 

\subsubsection{Watermarking}

A paper \cite{AI-Auth-Conf051} 
have showed that deep learning methods accompanied by dynamic watermarking imposes high accuracy of detecting cyber attacks such as eavesdropping, and man-in-the-middle.

\subsubsection{Secret Sharing}

The (n,n) secret sharing scheme lets the authentication process can be performed due to different situations. If all users are group members, they will pass the authentication. If not, the scheme is able to preprocess and apply authentication before and it can recognize the non-members\cite{AI-Auth-Conf167}.

\subsubsection{Medical Technologies}

There is a huge amount of medical data generated everyday. Those data are require to be protected against unwanted disclosure. AI-assisted authentication can achieve this goal efficiently.

Different medical technologies have been utilized for AI-based authentication purposes. Some of these technologies and related research works are reviewed in the following.

\begin{itemize}

	\item \textbf{Electrocardiograph:}

The electrocardiograph (ECG) is the most suitable way to authenticate patients. Some of them have limitations on body movement, eyelids open or speaking and hearing. Authentications using ECG can easily apply to almost everyone. A machine learning-based ECG authentication method \cite{AI-Auth-Jour035} is proposed. Researchers soon found out using RR-Interval Framed ECG will help to improve authentication \cite{AI-Auth-Jour040}.
	
The devices used to perform such authentication need to set their power consumption as low as possible. The work introduced in \cite{AI-Auth-Jour049} presents a hardware design to support authentication. After that, researchers embedded the neural networks into the design \cite{AI-Auth-Conf171}.
	
To boost the accuracy and efficiency of ECG authentication, deep learning start to be involved \cite{AI-Auth-Conf073}. With the help of ANN \cite{AI-Auth-Conf123}, 1D-CNN \cite{AI-Auth-Conf168} and End-to-End CNN \cite{AI-Auth-Conf156}, the accuracy and speed of the ECG-based authentication were significantly lifted.

	\item \textbf{Electroencephalograph (EEG):}

Same as the ECG, the Electroencephalograph (EEG) is the common biometrics for everyone. Hence, it is suitable for authentication.
	
An EEG-based deep learning authentication model has been provided \cite{AI-Auth-Conf057}. This model can continuously authenticate a person through the graph. Soon after the model above, a visual stimulation technology has been applied to the model \cite{AI-Auth-Conf086}. As CNNs can process images very efficiently with high accuracy, A CNN-based EEG authentication scheme was presented in \cite{AI-Auth-Conf150}. Wavelet transform can provide detailed edge features of a graph or image. Hence, a Wavelet transform-based machine learning authentication method using invisible visual stimulated EGG came out \cite{AI-Auth-Conf046}.

	\item \textbf{Ballistocardiograph:}

allistocardiograph (BCG) can be used to authentication a person as well. A RNN-based BCG authentication method \cite{AI-Auth-Conf170} is introduced. The accuracy of BCG is not as high as that of ECG, but when we combine those two graphs, the accuracy can achieve 100\%.

	\item \textbf{Electromyography (EMG):}

Electromyography (EMG) combined with deep learning methods can authentication as well. For example, \cite{AI-Auth-Conf125} presented an ANN-based EMG authentication scheme respectively. The one in \cite{AI-Auth-Conf120} can reach an accuracy of 81.6\% while the other one provided an accuracy of 95.0\%.

	\item \textbf{Radiograph:}

Hand Radiograph can be used for authentication a person, since it can provide a detailed pattern of hand bones. Joshi et al \cite{AI-Auth-Jour013} presented a method based on deep learning algorithms for forensic use. Moniga et al \cite{AI-Auth-Conf112} used ANN to process X-ray graphs, which helps to surge up the accuracy.

\end{itemize}

\subsection{Design Goals and Objectives}

\subsubsection{Performance}

A research paper\cite{AI-Auth-Conf109} 
 suggested that Principal Component Analysis (PCA) provides the best performance while reducing the input data dimensions. Another study \cite{AI-Auth-Conf130}  showed how the performance of authentication is effected by the size of neural network size. Another research \cite{AI-Auth-Conf166} have also concluded, calculating the time interval between keystrokes provides better performance comapred to other authentication methods.

\subsubsection{Robustness}

For identity authentication, security and privacy are difficult to balance. Hence, Gu et al \cite{AI-Auth-Jour008} proposed a secure and privacy-preserving scheme for Wi-Fi authentication based on the behavior. The main idea is to used deep learning method to learn from the client behavior and decide if a client is legitimate.

\subsubsection{Timeliness}

The authors of \cite{AI-Auth-Conf124} have proposed a new real time face authentication method that provides less than 1\% error in real time authentication. In a similar research\cite{AI-Auth-Conf131} have used neural networks to authenticate users in real time.

\subsubsection{Cost}

Low consumption and high efficiency are essential features of the certification system, especially for continuous authentication algorithms \cite{a18}. Researchers designed a smartphone-based low cost authentication scheme combining with multiple biometrics captured by sensors \cite{AI-Auth-Conf163}. According to their experiment, the proposed scheme improved the accuracy and achieved a real-time authentication with the time consumption of 0.032 second.

\subsubsection{Tradeoffs}

A research\cite{AI-Auth-Conf031} 
 that was conducted in 2017 evaluated strength, and performance of the Biometic Authentication Systems (BAS).

\section{The Taxonomy}\label{Tax}

In this section, we discuss the taxonomy of AI-assisted authentication. We classified the methods by the type of AI models. Subsection "Learning" introduced the schemes based on general machine learning, while subsection "Neural Networks" reviewed the methods that leveraged different kinds of neural networks.

\subsection{Learning}

Many researchers have been focusing on the application of learning-based methods in authentication \cite{AI-Auth-Jour023}. This combination has shown remarkable success in avoiding attacks on bypassing systems\cite{AI-Auth-Jour024}. Learning-based methods have been used to authenticate users \cite{AI-Auth-Conf050}, different kinds of contents \cite{AI-Auth-Conf054}, etc \cite{a8}.

Some researchers have studied the application of one-shot learning \cite{AI-Auth-Jour008}, ensemble learning \cite{AI-Auth-Jour006}, federated learning \cite{a19} and multi-task learning \cite{AI-Auth-Conf076}.  Furthermore, adversarial learning has been studied in different authentication methods\cite{}{AI-Auth-Conf098} including biometric authentication \cite{AI-Auth-Jour005}. Researches have used biometric authentication and different environments such as IoT \cite{AI-Auth-Jour001} for Adversarial learning. Especially, additive adversarial learning \cite{AI-Auth-Jour027}has been studied in a few relevant research works \cite{AI-Auth-Conf087}.

A few researchers have examined supervised learning \cite{AI-Auth-Conf059} and semi-supervised learning \cite{AI-Auth-Conf062}\cite{AI-Auth-Conf089} for authentication purposes. However, reinforcement learning methods seem to be of more interest to the research community \cite{AI-Auth-Conf011}. They have been used in different environments including CANs \cite{AI-Auth-Jour021}\cite{AI-Auth-Conf060} and VANETs \cite{AI-Auth-Jour042}. Moreover, view-invariant \cite{AI-Auth-Conf058} and one-shot \cite{AI-Auth-Conf064} feature learning as well as adaptive metric learning \cite{AI-Auth-Conf015}  have been incorporated in different authentication schemes.

\subsubsection{Machine Learning}

The application of machine learning in authentication has been of interest to the research community \cite{a20}.

Researchers have used machine learning \cite{AI-Auth-Jour011} to analyze challenges and risks on mouse data.The authors of \cite{AI-Auth-Jour017} proposed a threshold-free PHY-AUC method that replaces the old methods and imposes a new method using classification based machine learning. Researches have also used the physical layer for fingerprint authentication\cite{AI-Auth-Jour018}\cite{AI-Auth-Conf028} \cite{AI-Auth-Conf030} and to prevent clone node and Sybil attacks\cite{AI-Auth-Jour022}. Offensive security researchers have proposed an attacking method \cite{AI-Auth-Jour025} that can efficiently steal users' credentials. The authors of 	\cite{AI-Auth-Jour026} also rely on keystrokes to authenticate users, the force and touch duration are extracted using the piezoelectric force touch panel. In \cite{AI-Auth-Jour033}\cite{AI-Auth-Jour035} 
 using electrocardiogram(ECG)-based biometric information
 An authentication method has been proposed. Researches \cite{AI-Auth-Jour040} have also used RR-Interval framed ECG for machine learning biometric authentication. Machine learning was also used in automobiles to authenticate drivers\cite{AI-Auth-Jour041}. 
  Machine learning was also used in many other fields, some models were implemented for channel based message authentication\cite{AI-Auth-Conf006}, Identification of Obstructive Sleep Apnea (OSA) patients\cite{AI-Auth-Conf007}, Securing remote connection by authenticating users using machine learning technique\cite{AI-Auth-Conf008} 
 and bank note authentication \cite{AI-Auth-Conf021}. Machine learning used CAPTCHA to verify smart-TV users by running a model that distinguishes between real human images and downloaded images to limit attacks on smart-TV and other IOT devices\cite{AI-Auth-Conf022}.Performance of authentication systems was also analyzed by machine learning\cite{AI-Auth-Conf031}. Markov models were used with machine learning to authenticate users \cite{AI-Auth-Conf032}. The authors of \cite{AI-Auth-Conf036} proposed using machine learning for risk based authentication.
 
 Machine learning was also used for Physical layer authentication\cite{AI-Auth-Conf069}\cite{AI-Auth-Conf093} \cite{AI-Auth-Conf091}, data compression physical layer authentication \cite{AI-Auth-Conf081} and Blind physical layer authentication \cite{AI-Auth-Conf101}. Human analysis\cite{AI-Auth-Conf038}. In \cite{AI-Auth-Conf046} electroencephalograms users were authenticated using machine learning. In another research \cite{AI-Auth-Conf047}  one-class nearest neighbor (OCNN) was proven to provide better security performance over statistical methods. Ahmadi et al\cite{AI-Auth-Conf055} proposed a method to authenticate a person by classifying the motion records such as walking, standing, sitting, going upstairs and downstairs using machine learning. In \cite{AI-Auth-Conf068}, researchers proposed a machine learning-based IPS to monitor the traffic and alert when a DDoS attack could be performed. In \cite{AI-Auth-Conf071}, a taxonomy was provided for IoT devices authentication using machine learning.
 
According to recent researches, we can try to detect misuse by using biometrics and machine learning.\cite{AI-Auth-Conf099} 
\cite{AI-Auth-Conf092}. Machine learning is used in many different authentication methods, some of these methods are users using mouse gestures\cite{AI-Auth-Conf087}\cite{AI-Auth-Conf084}. Gabor Filter and Support vector machine (SVM) \cite{AI-Auth-Conf096}, smartphones continous authentication\cite{AI-Auth-Conf085}, most frequently user region based\cite{AI-Auth-Conf084}, Human body odor\cite{AI-Auth-Conf072}, authorizing edge devices\cite{AI-Auth-Conf071} 
, and analyzing geometrics in biometric authentication systems\cite{AI-Auth-Conf080}.

\textcolor{black}{Different Types of machine learning methods have been used for authentication purposes. These types are discussed in the following.}

\paragraph{Incremental Learning}\mbox{}

Incremental learning is the process of taking new inputs and updating existing data according to the new data. In a recent paper \cite{AI-Auth-Conf063} researches have implemented incremental learning to IOT devices, irregular clustering was used to increase trust and the machine learning confidence as more data is inputted to analyze and detect new devices and authenticate them.

\paragraph{Transfer Learning}\mbox{}

Transfer learning is a subset of machine learning that has been trained to learn from a specific dataset to provide prediction or classification for a target problem. What's more, it can be used to solve a different but similar problem. This kind of learning can be achieved either by giving the algorithm a fine-tuning or by the freezing layer\cite{AI-Auth-Jour002}. By combining with Blockchain, a transfer learning-based authentication mathod for IoT devices has been presented in \cite{AI-Auth-Jour014}.

\paragraph{Federated Learning}\mbox{}

Federated learning is the process of running a machine learning model on private device and keeping the data private on individual devices \cite{AI-Auth-Conf090}.

\paragraph{Learning Automata}\mbox{}

Learning automata is a set of decision-making models that can learn from their previous status and situations. They are able to learn from the former experiences to optimize actions and direct a new work\cite{AI-Auth-Conf088}.

\paragraph{Few-Shot Meta Learning}\mbox{}

Few-shot meta learning is used when there is a small training data set to learn from. This method has very good learning performance given the fact that it takes a small training data size and is used \cite{AI-Auth-Conf014} for facial authentication without the need of a large data set.

\paragraph{In-Situ Machine Learning}\mbox{}

In-situ means "the original place".  In-situ machine learning describes a kind of machine learning model that does not move data around\cite{AI-Auth-Conf082}. The key of this design is to figure out a way to minimize the need of moving data in and out of databases. There are some benefits of machine learning, such as reducing the cost of time and storage space when processing data.

\subsubsection{Deep Learning}

\textcolor{black}{The application of deep learning in authentication has received a research focus in recent years.} Deep learning has been favored by researchers since its emergence, due to its efficiency and high accuracy \cite{a21}. Deep learning-based authentication methods have been a focus in recent years.

Deep learning always performs excellently in the biometrics authentication fields. For example, biometrics authentication approaches such as palm vein authentication in \cite{AI-Auth-Jour010}, hand radiographs authentication in \cite{AI-Auth-Jour013}, iris \cite{AI-Auth-Conf033} combined with Sclera authentication in \cite{AI-Auth-Conf012}, eye and mouth movement authentication \cite{AI-Auth-Conf053}, face\cite{AI-Auth-Conf065} and voice\cite{AI-Auth-Conf100} recognition authentication are using deep learning methods. For multiple biometrics authentication systems like \cite{AI-Auth-Conf020}, deep learning is always the first choice because it can accept input in different kinds of formats.

In some scenarios that required fast and secure authentication processing like smartphone authentication\cite{AI-Auth-Jour034}, deep learning n provide efficient continuous authentication solutions at a high accuracy level.  In \cite{AI-Auth-Conf074}, researchers made a detailed comparison to show the advantages of deep learning methods in mobile phone authentication.

In the IoT devices and network nodes authentication, some methods are based on deep learning \cite{AI-Auth-Jour037}. A dynamic digital watermarking identification using deep learning has been presented for IoT devices \cite{AI-Auth-Conf051}. As to transmitter authentication, \cite{AI-Auth-Conf052} provided a physical layer multiple transmitter authentication method that uses deep learning.

What's more, deep learning algorithms are popular in other subfields of authentication. For medical authentication, several pieces of research for deep learning-based methods using ECG \cite{AI-Auth-Conf073}, EEG \cite{AI-Auth-Conf057}\cite{AI-Auth-Conf086} achieved high accuracy. Deep learning can also contribute to Wi-Fi traffic-based passive authentication\cite{AI-Auth-Jour019}, side-channel-based devices authentication \cite{AI-Auth-Conf048}. For object authentication, algorithms in \cite{AI-Auth-Conf023} \cite{AI-Auth-Conf061} have shown the advantages of deep learning. An attacking method towards AI-assisted authentication in \cite{AI-Auth-Conf013} is using deep learning to process data.

\paragraph{Deep Metric and Feature Learning}\mbox{}

Deep feature learning (Deep Representation Learning) learns information about data directly from raw data and trains the machine learning model\cite{AI-Auth-Conf034}.

Deep metric learning compares data by distance, the training data is used to learn the distance between data and classify them. The Euclidean distance is what measures the space between data. \cite{AI-Auth-Jour003} \cite{AI-Auth-Jour009}.

\paragraph{Deep Reinforcement Learning}\mbox{}

Deep Reinforcement Learning (DRL) is a machine learning method that makes decisions by trial and error. It can take a large number of inputs and make a decision without any manual guidance. In \cite{AI-Auth-Conf070}, researchers use it to accelerate processing.

\subsection{Neural Networks}

The application of neural networks in authentication has been of great interest to the AI and security research communities \cite{a22}. As the demand for machine learning gradually increases and the content of learning becomes more complex, researchers are becoming more and more interested in neural networks.

\textcolor{black}{Neural networks were used to enhance the traditional authentication methods, such as password-based authentication \cite{AI-Auth-Jour048}, fuzzy authentication systems \cite{AI-Auth-Jour032}. Which are also used for biometrics authentication to provide high security.
In \cite{AI-Auth-Jour046} authors went over a paper from 2001 and showed how it is vulnerable to attacks using neural networks.
Neural networks are good at processing complex data. In \cite{AI-Auth-Conf130} the authors compared the used of 1 step with neural network of two different sizes, 9-9-1 and 16-16-1 layers and achieved a high result of 99.852\% and 99.985\%, respectively.Neural network was used for biometric authentication in \cite{AI-Auth-Conf078} for e-learning platforms. It was also used alongside DLT algorithm for authenticating 3D faces \cite{AI-Auth-Conf108}. Neural network was used for color facial authentication\cite{AI-Auth-Conf121}. Researchers in \cite{AI-Auth-Conf153} proposed a very interesting method of authentication using finger knuckles using neural networks.
Neural networks were used extensively in behavioral authentication. The  authors of \cite{AI-Auth-Conf133} used neural network to authenticate users by using cluster analysis of the time period between keyboard keystrokes. A similar research was conducted by \cite{AI-Auth-Conf127} to used the time latency between keystrokes to authenticate users with the help of neural networks and k-nearest neighbor algorithm. The use of keystroke with authenticating users without logging in was proposed in \cite{AI-Auth-Conf106}, the authentication of users is done as soon as the user starts typing and it provided great results as it only resulted in 1.1\% of false alarm rate. The use of keystroke authentication was compared in \cite{AI-Auth-Conf139}. 
Mouse behaviour authentication such as mouse movements, time between click was a proposed method by \cite{AI-Auth-Conf141}. Researchers in \cite{AI-Auth-Conf104} use neural network for faster authentication of users without having to input any passwords, but simply by confirming the reference value of each user and recognizing the user instantly. Offline handwritten signature authentication with neural network was used by \cite{AI-Auth-Conf114}.}

Neural networks usually can perfectly extract the feature from raw data.\cite{AI-Auth-Conf138} Generally, user behavior data are hard to process by regular machine learning, due to multiple dimensions of data. On some occasions like \cite{AI-Auth-Conf109} shows, researchers can use a neural network to reduce the complexity of data so that they can extract the features easily.


For health care data authentication, neural networks can accurately learn from very tiny details of the dataset and provide a prediction. For instance, \cite{AI-Auth-Jour049}\cite{AI-Auth-Conf171}\cite{AI-Auth-Conf123}\cite{AI-Auth-Conf140} presented methods for authentication using ECG, \cite{AI-Auth-Conf120}\cite{AI-Auth-Conf125} presented schemes for EMG-based authentication. Neural networks application in those approaches helped accurately deal with the raw data to get a better result.

In other fields like IoT, neural networks contributed to the emitter identification \cite{AI-Auth-Jour053}, as well as IC authentication on the physical layer \cite{AI-Auth-Conf002}. For group authentication, \cite{AI-Auth-Conf167} provided a method to secret share information. Integrated with Shannon's Information Theory, a multimode authentication scheme has been proposed \cite{AI-Auth-Conf134}.

\subsubsection{Probabilistic Neural Networks}

Probabilistic Neural Network is a feedforawrd neural network, operation in the neural network are organized into multi layers Input layer, Pattern layer, Summation layer, Output layer \cite{AI-Auth-Conf102}.

\subsubsection{Deep Neural network}

Deep neural network(DNN) is a specific type of deep learning that uses multiple layers of neural networks. Those networks can learn from the training data and do not need human work. The DNN can make decisions as a human does.

Deep neural networks are widely used for AI-based authentication. \cite{AI-Auth-Jour051} introduced a channel state information-based authentication method using deep neural networks. researchers have tried different neural networks like CNN and RNN. According to their experiment, they found out that the combination of CNN and RNN called CRNN performed the best. In \cite{AI-Auth-Conf009}, deep neural networks were used for facial recognition. CNN does well in image processing. Similarly, \cite{AI-Auth-Conf163} provided a herbal leaf recognition method using DNN.

For smartphone biometrics authentication, DNN can help to extract features from complex datasets efficiently and accurately. In \cite{AI-Auth-Conf111} \cite{AI-Auth-Conf149}, two different mobile phone authentication methods were proposed. \cite{AI-Auth-Conf163} presented a low energy cost smartphone biometrics authentication scheme.

For medical data authentication, Graphs like ECG with DNN were commonly used. \cite{AI-Auth-Conf154} presented two different schemes for medical data-based authentication using DNN.

\paragraph{Convolutional Neural Networks}\mbox{}

The convolution neural networks (CNN) are a set of means that contain multiple neural network layers based on convolution computing. This type of neural network is widely used in the image processing and computer vision fields. A 2018 research paper suggested using convolutional neural network where biometric information is infused together from different resources\cite{AI-Auth-Jour055}, another research proposed image hashing for authentication based on convolutional neural network\cite{AI-Auth-Jour043}.
Many research on authentication involving convolutional neural network has been doe,Hand motion authentication\cite{AI-Auth-Conf132}\cite{AI-Auth-Conf168} , Human authentication using ankle movement\cite{AI-Auth-Conf145}, continuous authentication for vhiecles\cite{AI-Auth-Conf142} . In \cite{AI-Auth-Conf112} 
\cite{AI-Auth-Conf010} authors used deep learning authentication, according to the research that has been conducted in 2018 \cite{AI-Auth-Conf103}, it is easy to achieve authentication by either hand motion or EMG using convolutional neural archittecture. In another study authors provided physical layer authentication using convolutional neural networks\cite{AI-Auth-Conf119}. The Triple-Pool CNN (TP-CNN) has been proposed in \cite{AI-Auth-Conf158} that uses Convolutional Neural Network. The Two-Stream CNN was designed in \cite{AI-Auth-Conf157}.  In \cite{AI-Auth-Conf161} authors proposed using facial attributes such as hair colour, gender et cetera to authenticate users this reduces the complexity of the network.

\textcolor{black}{The following types of convolutional neural networks has been used for different authentication purposes.}

\begin{itemize}
	\item 1-D, 2-D and 3-D Convolutional Neural Networks

	The 1D convolutional neural network is the simplest CNN that has only one convolution layer. In \cite{AI-Auth-Conf168}, 1D CNN has been used for pattern learning. It only contains one convolution layer and one max pooling layer for obtaining efficiency.

	\textcolor{black}{The 2D CNN emerged for improving the weakness of the 1D CNN.\cite{AI-Auth-Conf056} As for the sample size, the 1D CNN can only accept a few thousand. The 2D CNN can take 2D data as inputs like an image.}

	The 3D CNN can accept more inputs than the 2D version at a time. By using it, we can train the AI model with a lot more bunches of data. In \cite{AI-Auth-Conf135}, researchers use it to extract features from lip motion and voice samples. It highly accelerated the processing and bound the visual information and sound information together.

	\item Hierarchical Convolutional Neural Networks

	A hierarchical Convolutional Neural Network(HCNN) is similar to CNN that embedded a series of hierarchies to distinguish the easy classes and the difficult classes	\cite{AI-Auth-Conf128}. An HCNN will learn and classify the data by the difficulty label of the data.

	\item End-to-End Convolutional Neural Networks

An End-to-End (E2E) CNN is a specific way that people trains a CNN model and bypass the traditional layers in the middle	\cite{AI-Auth-Conf156}.

	\item Deep Convolutional Neural Networks

	A Deep Convolutional Neural Network (DCNN) is a combination of traditional CNN and Deep Learning. Usually, DCNN is used for image and video patterns detection. It can accept 3-D inputs and uses CNN as the middle layer of Deep Learning \cite{AI-Auth-Conf118}.

	\item Two-Stream Convolutional Neural Networks

	The Two-Stream CNN was designed in \cite{AI-Auth-Conf157}. They used both time-domain data and frequency-domain data as the inputs of a CNN to let it authenticate in the two fields.

	\item Triple-Pool Convolutional Neural Networks

	The Triple-Pool CNN (TP-CNN) has been proposed in \cite{AI-Auth-Conf158}. A TP-CNN includes $n$ convolutional blocks that are allowed to connect with a hidden layer. It is distinguished from the traditional CNN in an exponential linear unit.

\end{itemize}

\subsubsection{Adaptive Neural Networks}

Adaptive Neural Network is a group of networks in a rapidly changing environment. In \cite{AI-Auth-Jour039} adaptive neural network was used to detect changes made to a network for physical layer authentication.

\subsubsection{Bayesian Neural Networks}

The Bayesian Neural Network (BNN) is a mixture of the Bayesian framework and the DNN. It does not provide the point estimation, but it contributes to expressing the estimate and an uncertainty range.\cite{AI-Auth-Conf005}.

 s

\subsubsection{Symbiotic Neural Networks}

Symbiotic Neural Networks used in \cite{AI-Auth-Conf148} used to analyze psychological and hand gesture movements for a fast hand gesture authentication.

\subsubsection{Siamese Neural Networks}

The Siamese Neural Network (SNN) is a learning scheme that uses the same wights to process two separated inputs and output two streams of comparable data\cite{AI-Auth-Conf151}. With the help of SNN, an online authentication system based on handwriting recognition has been presented in \cite{AI-Auth-Conf110}.

\subsubsection{Back Propagation Neural Networks}

Back propagation is the process of iterating backwards to a neural network to minimize the error value. Back propagation neural networks were used in \cite{AI-Auth-Conf146} to authenticate offline signatures.

\subsubsection{Counter-Propagation Neural Networks}

The Counter-propagation Neural Network is used for embedding copyright information in \cite{AI-Auth-Conf115}. Researchers embedded information into Full Counter-propagation Neural Network(FCNN) layer instead of into cover image to make the processing easier. Besides, the FCNN layer can contain more information than the cover image.

\subsubsection{Flexible Neural Networks}

Flexible Neural Networks have different layers that are made of neurons, each layer is matched up with other neurons based on similar set features\cite{AI-Auth-Conf162}.

\subsubsection{Weightless Neural Networks}

The Weightless Neural Network(WNN) is similar to Artificial Neural Network but it only requires one-pass learning.\cite{AI-Auth-Conf137} It uses RAM to solve the image pattern recognition problem. Because it is based on memory, the processing speed is very high which makes the problem very simple.

\subsubsection{Feed Forward and Recurrent Neural Networks}

\paragraph{Feed Forward Recurrent Neural Networks}\mbox{}

The following types of feed forward neural networks have been considered in recent research on the applications of neural networks in authentication systems.

\begin{itemize}

	\item  Extreme Learning Machine
	In \cite{AI-Auth-Jour012}\cite{AI-Auth-Jour023} authors have argued that threshold in hypothesis is not always reliable. Therefore; they proposed a new physical-layer to authenticate dynamic-networks using extreme learning machine. Another study \cite{AI-Auth-Conf019} uses the extreme learning machine generate a robust digital watermark that can resist multiple types of attack.

	\item  Multi-Layer Perceptron (MLP) Neural Networks

	Neural network was used 
	\cite{AI-Auth-Conf122} to authenticate keystrokes by investigating the performance of keystroke dynamics and using multilayer perceptron.

	\item Radial Basis Function (RBF) Neural Networks

	Radial Basis Function Neural network was used \cite{AI-Auth-Conf116} 
     for face authentication.

\end{itemize}

\paragraph{Recurrent Neural Networks}\mbox{}

Recurrent Neural Networks(RNN) is one of the most popular deep neural networks for authentication. It can be used for improving traditional signature authentication \cite{AI-Auth-Conf042}. It can also be applied for physical layer authentication, such as transmitter authentication and channel prediction for IoT networks \cite{AI-Auth-Jour050}, and mobile authentication based on the channel state information \cite{AI-Auth-Conf165}.

For health care data authentication, RNN can help deal with the dataset, and significantly lift the accuracy of systems. Based on the ECG information, RNN is able to extract useful features to distinguish different people\cite{AI-Auth-Conf160}\cite{AI-Auth-Conf170,bbb}.

Among different types of recurrent neural networks, the following ones seem to be of more interest to researchers.

\begin{itemize}

	\item \textbf{Hopfield Neural Networks:}

	A Hopfield Neural Network has one layer of $n$ recurrent neurons. It was introduced by Hopfield to optimize the tasks. Usually, it can accept multiple types of input and give one output.
	
	Traditional password-based authentication was proved unsecured.  \cite{AI-Auth-Jour044} and \cite{AI-Auth-Conf105} provided two different approach to enhance the password-based authentication by using HNN to make password registration and changing process more efficient and easier.

	\item \textbf{Long Short-Term Memory (LSTM) Neural Networks:}

	An LSTM Neural Network is a Recurrent Neural Network(RNN) with Long Short-Term Memory. The goal is to feed data directly to the NN. LSTM units can collect the temporal dynamics which is a necessity for a man to make a recognition based on multiple scenarios \cite{AI-Auth-Conf159}.

	\item \textbf{Autoassociative Neural Networks:}

	The Associative Neural Networks(NN) are a bunch of NNs that simulate Short-and-Long-Term memory. An auto-associative NN has been used in \cite{AI-Auth-Conf124}, which contains five layers to extract the features like a gray level distribution from images.

\end{itemize}

\begin{figure}[htbp]
	\centerline{\includegraphics[width=0.5\linewidth]{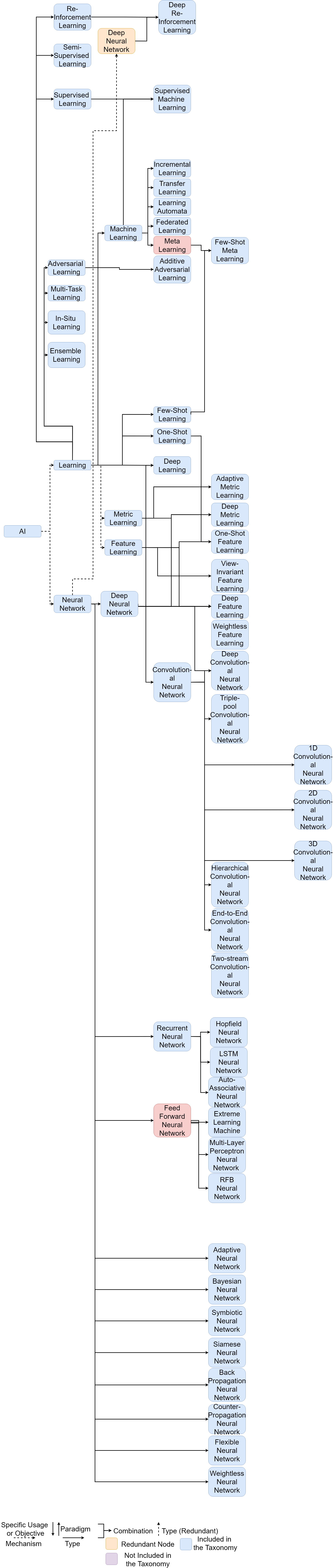}}
	\caption{AI-Assisted Authentication Taxonomy}
	\label{Fig-Taxonomy}
\end{figure}

\textcolor{black}{Figure \ref{Fig-Taxonomy}} shows the tree chart of taxonomy. It fully described all the categories mentioned above by providing the relationship between different kinds of machine learning methods. You can clearly see the connections between each classification, and all the subclasses.

\section{Future Roadmap}\label{Road}

We anticipate that research on AI-assisted authentication will move towards bio-inspired and quantum-inspired AI-assisted authentication in near future. Our reason for such an anticipation is the existence of trends discussed in Subsections \ref{sub1}, and \ref{sub2}.

\begin{figure}
    \centering
    \includegraphics[scale=0.16]{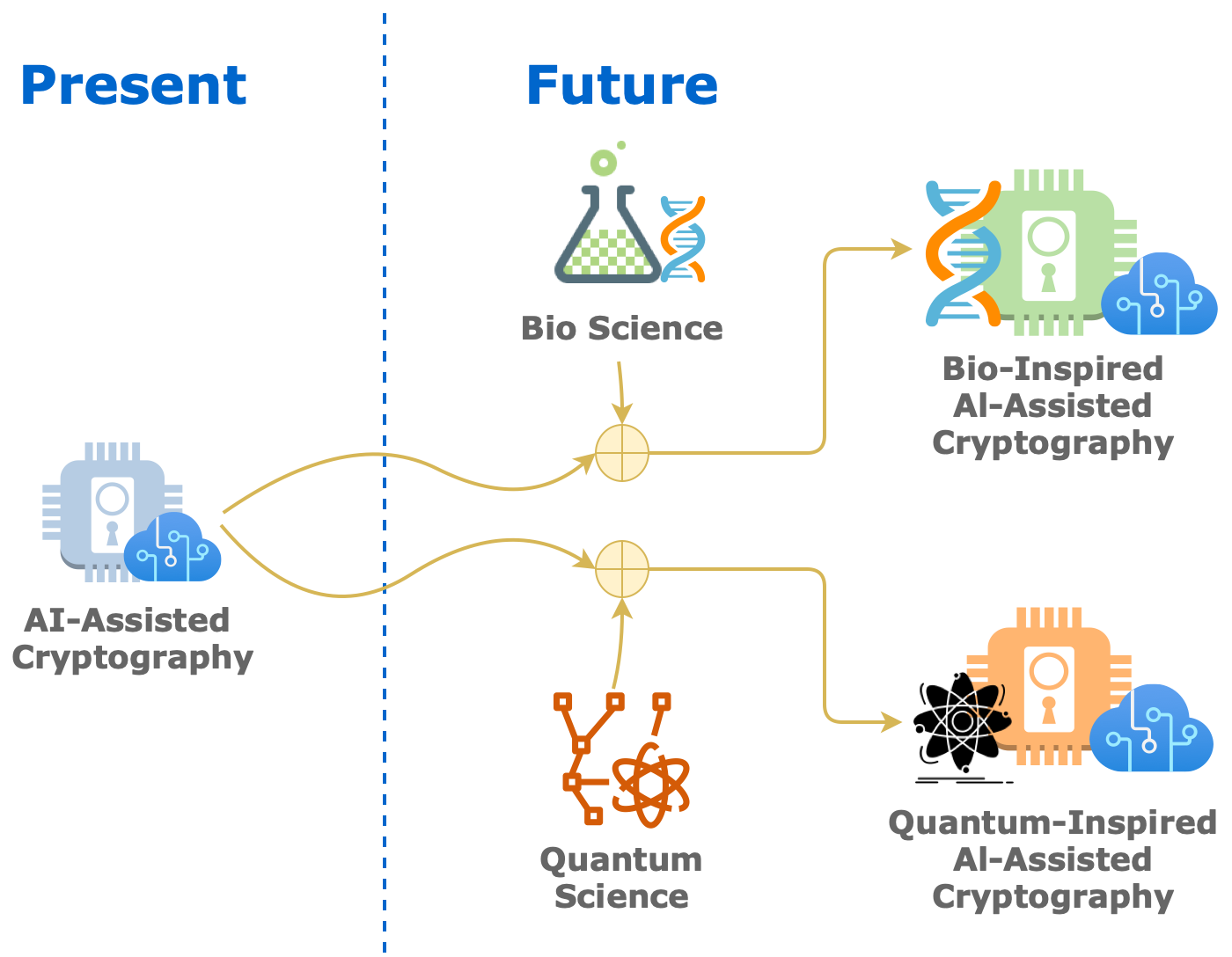}
    \caption{AI-Assisted Authentication Future Roadmap}
    \label{fig:Roadmap}
\end{figure}

\textcolor{black}{Figure \ref{fig:Roadmap} shows the trend of AI-assisted authentication combined with Bio-science and Quantum science. The graph was divided into two parts: present and future. The upper flow indicates that AI-assisted authentication may be able to integrate with bio-science to produce bio-inspired AI-assisted authentication. The lower flow demonstrates how quantum science can inspire AI-assisted authentication to construct quantum-inspired AI-assisted authentication. In the future, Bio-Inspired learning will have a profound impact on authentication. Bio-Inspired and Quantum-Inspired authentication will be leveraged to further improve the accuracy and usability of authentication.}

\subsection{Quantum-Inspired AI}\label{sub1}

Quantum-Inspired AI is the new trend of artificial intelligence. Quantam-inspired AI can offer a balance between exploration and exploitation \cite{Quant-AI-Jour001}. Other researchers\cite{Quant-AI-Jour002}
 have also concluded that the use of Deep Reinforcement Learning with quantam inspire AI offers a balance between exploration and exploitation. Robust Quantum-Inspired Reinforcement Learning proposed by\cite{Quant-AI-Jour004}
was used for real mobile robots. The authors of \cite{Quant-AI-Jour003} quantum-inspired multidirectional associative memory (QMAM) introduced with one shot learning. Researchers proposed a Quantum-inspired Fuzzy Based Neural Network (Q-FNN) \cite{Quant-AI-Jour005} that provides very high accuracy and sensitivity.

\subsection{Bio-Inspired Learning}\label{sub2}

Bio-Inspired machine learning is a future machine learning scheme inspired by biological structures.  This special category of machine learning algorithms could help deal with some authentication related to computer vision. So far, we have applied Bio-Inspired learning into some AI fields, such as visual attention prediction \cite{Fut-Bio-Jour001}, facial aesthetic prediction \cite{Fut-Bio-Jour002}. Besides, Bio-Inspired learning can be used for spam emails detection \cite{Fut-Bio-Jour004}, autonomous navigation \cite{Fut-Bio-Jour005} and UAV path planning \cite{Fut-Bio-Jour007}. For IoT, researchers have implemented it into multiobject optimization \cite{Fut-Bio-Jour006}. In the robotic field, Bio-Inspired learning contributes to simulating the action of animals to provide a guide for robots development \cite{Fut-Bio-Jour003}.

In conclusion, we have reviewed multiple research papers that have improved the authentication of users dramatically. In this review paper we have started with defining the most commonly used terms in artificial intelligence (AI) and machine learning, followed by that we have reviewed multiple research papers on how machine learning and authentication is combined to result in AI assisted authentication and the different methods used such as Back Propagation Neural Networks and Symbiotic Neural Networks, and different methods of authentication.

\section{Conclusions and Further Works}\label{Conc}

\textcolor{black}{This paper provides a comprehensive review of authentication, specifically focusing on the use of newly proposed authentication using Artificial Intelligence (AI). The introduction of commonly used terms, background knowledge about AI and authentication. The review of existing surveys, the state of art, taxonomy and future road map for AI-Assisted authentication. The use of AI has expanded the way researchers authenticate users and to pave a path to work on advance methods of authentication. This work provides researchers with an understanding on what problems to tackle in the future regarding authentication, and to present newer methods for AI Assisted authentication.
}

\bibliographystyle{IEEEtran}
\bibliography{References}

\end{document}